\documentclass{camera}

\newcommand{\psim}{\lower.5ex\hbox{$\; \buildrel \propto \over\sim \;$}}
\newcommand{\lesssim}{\lower.5ex\hbox{$\; \buildrel < \over\sim \;$}}
\newcommand{\gtrsim}{\lower.5ex\hbox{$\; \buildrel > \over\sim \;$}}

\newcommand{\e}{\epsilon}

\begin{document}

%
\title{GRBs as ultra-high energy cosmic ray sources: clues from {\it Fermi }}

%
\author{C.\ Dermer}

%
\organization{Naval Research Laboratory, Code 7653, 4555 Overlook Avenue, SW\\
Washington, DC 20375-5352 USA}

\maketitle

\begin{abstract}
If gamma-ray bursts are sources of ultra-high energy cosmic rays, then 
radiative signatures of hadronic acceleration are expected in GRB 
data. Observations with the {\it Fermi Gamma-ray Space
Telescope (Fermi)} offer the best means to search for evidence of UHECRs in 
GRBs through electromagnetic channels. Various issues related to UHECR acceleration in GRBs are reviewed, 
with a focus on the question of energetics.
\end{abstract}

\section{Introduction}

{\it Fermi } observations of GRBs provide a new
probe of particle acceleration in the relativistic outflows of GRBs.
Some generic features of the high-energy behavior of 10 Large Area Telescope (LAT) GRBs 
consisting of 8 long-soft and 2 short-hard GRBs at the time of this conference have been 
identified,
as summarized by Omodei \cite{omo09}. 
These are:
\begin{enumerate}
\item A delayed onset of the  $\gtrsim 100$ MeV radiation 
observed with the LAT compared with the start of the keV/MeV GBM radiation;
\item Long-lived LAT emission extending well after the Gamma ray Burst Monitor (GBM) radiation has fallen
below background, as known previously for long duration GRBs from EGRET \cite{hur94};
\item Existence of a distinct hard spectral component 
in addition to a component described by the Band function in both long and short GRBs, 
confirming the discovery of such a component from  joint BATSE/EGRET TASC analyses \cite{gon03}.
\end{enumerate}

In this contribution, the possibility that these features can be explained by 
UHECRs in GRBs is considered. Because protons and ions are weakly radiative 
compared to electrons, even with escaping energies $E \approx 10^{20}$ eV needed 
to explain UHECRs, large amounts of nonthermal
hadronic energy are required. We consider energetics arguments to constrain models
for UHECRs from GRBs.

\section{Maximum energy release in GRBs}

Bohdan Paczy\'nski, in his seminal article on hypernovae  \cite{pac98}, proposed
an explanation for the large apparent isotropic energies of GRBs 
by appealing to
the enormous energy available in 
the process whereby the core of a massive star collapses to form a long GRB. 
He noted that the $\sim 10$ M$_\odot$ core of a Type II supernovae carries $\approx 5\times 10^{54}$ 
erg of rotational energy, which could be tapped through Blandford-Znajek processes if surrounded
by a highly magnetized torus formed during the stellar collapse event. 
Paczy\'nski's arguments suggest that the maximum energy available from core collapse supernovae is 
therefore 
\begin{equation}
{\cal E}_{max}\lesssim 10^{54}~{\rm  erg}\;,
\label{calE}
\end{equation}
with perhaps an order-of-magnitude less in the coalescence events thought to form the short hard GRBs.

The association of long-duration GRBs with Type Ib/c rather than Type II 
supernovae does not alter his energetics argument, but the recognition that GRBs are jetted greatly reduces the 
absolute energy requirements.  Energy extracted through Blandford-Znajek processes is 
likely beamed in view of the rapid rotation of the newly formed black hole. Even so, we can still
apply the energy bound given by eq.~(\ref{calE}) to determine the plausibility of models with 
large energy requirements.

\section{Energetics of UHECRs from GRBs}

The idea that GRBs could accelerate UHECRs was proposed by 
Waxman \cite{wax95} and Vietri \cite{vie95} in 1995, 
and subject to criticism on energetics grounds \cite{ste00}, though
here related to whether GRBs within the GZK radius have the requisite volume- and time-averaged luminosity density. 
We can reformulate their argument in light of new data. 
The Auger results show that the differential energy density of UHECRs 
at $10^{20}$ eV is $\approx 2\times 10^{-22}$ erg cm$^{-3}$. 
The horizon distance
for $ 10^{20}$ eV UHECR protons is $\approx 50$ Mpc (shorter than the mean-free-path 
of $\approx 140$ Mpc, because this is the distance from which protons with measured 
energy $E$ originally had energy $\approx 2.7 E$), so that the required
emissivity to power $\gtrsim 10^{20}$ eV UHECRs
is $\dot\varepsilon_{CR}\approx c\times  2\times 10^{-22}$ erg cm$^{-3}$/50 Mpc 
$\approx 4\times 10^{43}$ erg Mpc$^{-3}$ yr$^{-1}$.  
If UHECRs are protons, then  $\dot\varepsilon_{CR}\approx 10^{44}\dot\varepsilon_{44}$ erg 
Mpc$^{-3}$ yr$^{-1}$, with $\dot\varepsilon_{44}\simeq 1$, is required to power $\gtrsim 10^{19}$ eV UHECRs, 
noting that the exponentially decreasing horizon distance with energy approximately cancels
the $\propto E^{-2}$ decrease in UHECR energy density over this energy range.

With long-duration GRBs taking place about twice per day over the full sky for BATSE-type
detection capabilities \cite{ban02}, and are found at a 
typical redshift of unity, implying a local 
space density for unbeamed sources of $\sim 2 \times 365{\rm~yr}^{-1}\zeta/[4\pi(4{\rm~Gpc})^3/3]\approx 0.3(\zeta/0.1)$
Gpc$^{-3}$ yr$^{-1}$, consistent with the value of $\approx 0.5$
Gpc$^{-3}$ yr$^{-1}$ found in more detailed treatments \cite{gpw05}. Here $\zeta$ accounts for 
the smaller star-formation activity occurring at $z\approx 1$ than at the present epoch \cite{yuk08}. 
If each GRB releases an amount of energy 
${\cal E}_{CR}$ in UHECRs, then the local volume- and time-averaged
cosmic-ray emissivity is  $\dot\varepsilon_{CR} \simeq 5\times 10^{-10} {\cal E}_{CR}(\zeta/0.1)$ Mpc$^{-3}$ yr$^{-1}$.
Equating this with the emissivity required to power the UHECRs implies that each GRB must release
${\cal E}_{CR}\approx 2\times 10^{53}\dot\varepsilon_{44}(\zeta/0.1)$ erg in UHECRs. This can be compared
with the typical electromagnetic energy release per GRB of $4\pi (4{\rm~Gpc})^2 10^{-5}{\cal F}_{-5}$ erg cm$^{-2}
\approx 2\times 10^{52}{\cal F}_{-5}$ erg, where ${\cal F}=10^{-5}{\cal F}_{-5}$ erg cm$^{-2}$ 
is the average BATSE long-duration GRB fluence.
Thus the energy released in UHECRs has to be $\gtrsim 10\times$ the energy measured from electromagnetic processes, 
independent of beaming.

The need for large baryon loads in GRB blast waves is confirmed
by detailed fits to the UHECR energy spectrum, which imply a factor 
$\approx 10$ -- 100 times more energy in  UHECRs than observed in  
electromagnetic radiation \cite{da06}. An explanation for the rapid
declines in {\it Swift} X-ray light curves in terms of escaping UHECR neutrons
from photohadronic production also requires highly baryon-loaded 
GRB outflows \cite{der07}.

\section{Minimum bulk Lorentz factor}

A $\delta$-function approximation for the $\gamma\gamma$ opacity constraint, given 
the detection of a $\gamma$-ray photon with energy $m_ec^2\e_1$ and variability time 
$t_v$, implies a minimum bulk Lorentz factor
\begin{equation}
\Gamma_{\rm min} \approx \left[ {\sigma_{\rm T} d_L^2 (1+z)^2 
f_{\hat \e}\e_1\over 
4t_v m_ec^4}\right]^{1/6}\;\;,\;\;
\hat \e = {2\Gamma^2\over (1+z)^2\e_1}\;.
\label{eq2}
\end{equation}
where $f_\e $ is the $\nu F_\nu$ flux at photon energy $m_ec^2 \epsilon$.
Thus $\Gamma_{min} \approx 916 [f_{-6} \e_1({\rm 3~GeV})/t_v({\rm s})]^{1/6}$, 
where the $\nu F_\nu$ flux is $10^{-6}f_{-6}$ erg cm$^{-2}$ s$^{-1}$, using 
values corresponding to time bin b for GRB 080916C \cite{abd09}. This GRB,
at redshift $z \cong 4.35\pm 0.15$ and luminosity distance $d_L = 1.25\times 10^{29}$ cm, 
had a total 10 keV -- 10 GeV energy fluence ${\cal F} = 2.4\times 10^{-4}$ erg cm$^{-2}$,  implying
a total $\gamma$-ray energy release ${\cal E}_{\gamma,iso} \cong 8.8\times 10^{54}$ erg.

Provided the target photon number spectrum 
has an index softer than $-1$, eq.~(\ref{eq2}) gives a result within $\sim 10$\% of a numerical 
integration over spectral parameters. Issues in the evaluation of the uncertainty, $\Delta\Gamma_{min}$, 
in the value of $\Gamma_{min}$ include
(1) uncertainties in redshift and $\nu F_\nu$ spectral flux; (2) the definition of 
variability time $t_v$; (3) the cospatial assumption that the target photons are made in the same
region as the high-energy photon; (4) the assumed geometry and dynamical state \cite{gra08} of the emission region, 
giving the escape probability of a high-energy photon; (5) statistical fluctuations for the 
detection of a high-energy photon that furthermore depend on the actual high-energy photon spectrum. 

Writing the GRB blast wave Lorentz factor $\Gamma = q \Gamma_{min}$, a consideration of these various
issues  show that $q \gtrsim 0.5$ represents a reasonably conservative expectation for the actual value of 
$\Gamma$. The energetics of a given model 
can depend strongly on $\Gamma$. 

\section{Emission radius: internal or external shocks?}

The total apparent energy release of GRB 080916C is
${\cal E}_{iso} = 10^{55} {\cal E}_{55}~{\rm erg}$, with ${\cal E}_{55}\gtrsim 1$.
 The corresponding deceleration length is 
$r_{dec} = 1.2\times 10^{17}{\cal E}_{55}^{1/3}/n\Gamma_3^{2/3}$ cm, where $10^3\Gamma_3$
is the coasting Lorentz factor and $n$(cm$^{-3}$) is the external medium density.
The implied radius for internal shock emission is 
$r \cong \Gamma^2 c t_v/(1+z) \cong 6\times 10^{15}\Gamma_3^2 t_v({\rm s})^2$ cm.
The unexpectedly large emission radius, close to the deceleration radius when $q\approx 2$, has
led a number of researchers to argue that the LAT radiation originates from leptonic
synchrotron radiation at an external forward shock \cite{kbd09,ghi09}. Another 
possibility is that the delayed LAT emission results from upscattered cocoon 
radiation \cite{twm09}.

\section{$\gamma$ rays from UHECRs in GRB blast waves}

We can make some simple estimates to deduce the total energy needed
to obtain bright hadronic $\gamma$-ray emission from GRBs 
through proton synchrotron and photopion processes \cite{rdf09, wan09,asa09}. 
Here we consider only UHECR proton acceleration, using parameters appropriate to GRB 080916C. 

The internal photon energy density $u^\prime_\gamma\cong d_L^2\Phi/R^2\Gamma^2 c$, 
where $\Phi = 10^{-5}\Phi_{-5}$ erg cm$^{-2}$ s$^{-1}$ is the measured energy flux,
$R\cong \Gamma^2 c\hat t/(1+z)$ is the shock radius, and $\hat t$ is a fiducial timescale (corresponding
to $t_v$ for internal shocks, or the GRB duration for an external shock). Writing the 
magnetic-field energy density $u^\prime_B = \epsilon_B u^\prime_\gamma$, where $u^\prime_B = B^2/8\pi$, 
then the magnetic field in the emission region of GRB 080916C is
 $B({\rm kG}) \cong 2.0 \sqrt{\epsilon_B \Phi_{-5}}/\Gamma_3^3 \hat t({\rm s})$.
The Hillas criterion whereby the Larmor radius $r^\prime_{\rm L}= m_pc^2 \gamma_p^\prime /eB < \Delta R^\prime \cong R/\Gamma$, 
where $\Delta R^\prime$ is the comoving shell width, 
implies that the escaping UHECR proton Lorentz factor $\gamma_p \cong \Gamma \gamma_p^\prime 
\lesssim ceB\Gamma^2 \hat t/(1+z)m_pc^2$. For GRB 080916C, this relation implies that UHECRs can be accelerated to 
$\gamma_p \lesssim (2d_L e/\Gamma m_pc^2)\sqrt{2\pi \epsilon_B\Phi/c} \cong 4\times  10^{12} \sqrt{\epsilon_B \Phi_{-5}}/\Gamma_3$,
independent of time (note that $B \propto t^{-1}$, with the divergence at $t\rightarrow 0$ 
prevented due to the time dependence of $\Phi$), or to 
energy $E_p \lesssim 2\times 10^{21} B({\rm kG}) \Gamma_3^2 \hat t(s)$ eV. 

A further restriction on the maximum proton energy is obtained by balancing 
the acceleration rate, given by $c/(\phi r^\prime_{\rm L})$, with $\phi \gg 1$, 
with the synchrotron loss rate. This gives $\gamma_{sat,p} \cong 2\times 10^{12} \Gamma_3/\sqrt{(\phi/10) B({\rm kG})}$, 
comparable to the value obtained above from the Hillas condition for the chosen parameters.

\subsection{Proton synchrotron energy requirements}

The proton synchrotron energy loss timescale, as measured by an observer,
is $t_{syn} \cong3m_ec^2 (1+z)/[4\Gamma \mu^3 c\sigma_{\rm T} u^\prime_B \gamma_p^\prime$], 
where $\mu \equiv m_e/m_p$. The typical photon energy (in $m_ec^2$ units) 
of the measured proton synchrotron emission is $\epsilon_{p,syn} \cong 
\Gamma\mu B\gamma_p^{\prime~2}/[(1+z)B_{cr}]$, where $B_{cr} = 4.414\times 10^{13}$ G
is the critical magnetic field. From this, one obtains the jet power associated
with the magnetic field, given by
\begin{equation}
L_B \cong \frac{ R^2 c\Gamma^2 B^{2}}{2} 
\cong {2\times 10^{58}\Gamma_3^{16/3} t^{2/3}_{syn}({\rm s})
\over E_\gamma({\rm 100 ~MeV})^{2/3} } \;
{\rm erg~s}^{-1}
\label{LB}
\end{equation}
\cite{wan09}. The absolute energy requirements are ${\cal E}_{abs} \cong 
L_B t_{syn}f_b/(1+z) \cong {1.6\times 10^{59}\Gamma_3^{16/3}f_b~ t^{5/3}_{syn}({\rm 10~ s})
/ E_\gamma({\rm 100~MeV})^{2/3} }$ erg, where $f_b$ is a beaming factor. 
In the scenario of Ref.\ \cite{rdf09}, the delayed onset corresponds to the time for protons to accumulate 
and cool such that they are radiating most of the proton-synchrotron photons near energy $E_\gamma$

Such large energies disfavor this interpretation for the LAT radiation. To salvage the proton 
synchrotron model \cite{rdf09}, a narrow jet opening angle of order $1^\circ$ along with a value of 
$q \cong 0.5$ gives ${\cal E}_{abs}\cong$
${4\times 10^{53}(\Gamma_3/0.5)^{16/3}(f_b/10^{-4})~ t^{5/3}_{syn}({\rm 10~ s})
/ E({\rm 100~MeV})^{2/3} }$ erg, within acceptable ranges. Another way to avoid the large energy losses is 
to suppose that the UHECR protons are accelerated to their allowed maximum energy and radiate proton 
synchrotron photons that cascade to energies $\lesssim 100$ MeV. This possibility seemed unlikely 
given that the spectrum of GRB 080916C is consistent with a single Band function \cite{abd09}. The detection of distinct
components in GRB spectra suggests that a cascading interpretation be more carefully considered \cite{asa09,asa09a}.

\subsection{Photohadronic energy requirements}

Proton-photon interactions making secondary pions, $\gamma$ rays, and neutrinos represents another
likely channel for making a $\gamma$-ray component identifiable in the {\it Fermi} data.  The efficiency for 
extracting the energy of a proton with escaping energy $E_p$ from photohadronic processes can be written as $\eta_{p\gamma}(E_p)
= t_{dyn}/t_{p\gamma}(E_p) \cong (R/\Gamma c) t^{-1}_{p\gamma}(E_p)$, where $t_{dyn}$ is the 
dynamical time scale, the rate for photohadronic
energy losses is $ t^{-1}_{p\gamma}(E_p) \cong c(K_{p\gamma}\sigma_{p\gamma}) \int_{\epsilon^\prime_{thr}}^\infty 
d\epsilon^\prime n^\prime(\epsilon^\prime)$, and the comoving photon spectrum 
$n^\prime(\epsilon^\prime) \cong d_L^2 f_\epsilon/(m_ec^3 \epsilon^{\prime~2} R^2\Gamma^2)$, 
with $\epsilon^\prime \cong (1+z)\epsilon/\Gamma $. Here  $K_{p\gamma}\sigma_{p\gamma} \cong 70\mu$b above
threshold photon energy defined by $\gamma^\prime \epsilon^\prime \cong 400$ \cite{da06}.  
Defining protons with energy $E_p^{pk}$ which interact at threshold with photons with energy
$\epsilon_{pk}$ at the peak $f_{\epsilon_{pk}}$ of the $\nu F_\nu$ spectrum, 
then $E_p^{pk} \cong 400 m_pc^2 \Gamma^2/[(1+z)\e_{pk} \cong 7\times 10^{16}\Gamma_3^2/\e_{pk}$ eV. 
One obtains 
\begin{equation}
\eta_{p\gamma}(E_p^{pk}) = { K_{p\gamma}\sigma_{p\gamma}d_L^2 f_{\epsilon_{pk}}\over 
\Gamma^4 m_ec^4 t_v(1-b) \epsilon_{pk}} \cong 0.015\; {f_{-6}\over \Gamma_3^4 t_v({\rm s})\e_{pk}}
\label{eta}
\end{equation} 
\cite{wax95,der02,mur09}. Here $b(<0)$ is the $\nu F_\nu$ index above $\e_{pk}$, 
so that $f_\epsilon = f_{\epsilon_{pk}}(\e/\e_{pk})^b$ when $\e > \e_{pk}$, and 
$\eta_{p\gamma}(E_p)\cong \eta_{p\gamma}(E_p^{pk}) (E_p/E_p^{pk})^{1-b}$. Likewise in the asymptotic 
limit $\e \ll \e_{pk}$, $\eta_{p\gamma}(E_p)\cong\eta_{p\gamma}(E_p^{pk}) (E_p/E_p^{pk})^{1-a}$, where $a$ is the $\nu F_\nu$ index
at energies below the peak energy (provided $a>1$).

The total efficiency for photohadronic production depends on the 
spectrum of accelerated protons and ions. If an $ E_p^{-2}$ spectrum is assumed, 
then the efficiency is reduced in proportion to the number of decades of weakly radiating
low-energy protons. This might not be too severe if the lowest energy 
protons have escaping energy $E_p\approx \Gamma^2 m_pc^2 \approx \Gamma_3^2$ PeV. 
Furthermore, nonlinear shock acceleration in colliding shells, and 
second-order processes in the shocked fluid can give harder cosmic-ray spectra, though
consistency when fitting the measured UHECR spectrum would constrain the assumed 
accelerated cosmic-ray spectrum.

The final expression in eq.\ (\ref{eta}), specific to parameters of GRB 080916C, shows that
$\sim 1$ -- 10\% of the  energy can be extracted via photohadronic processes by protons with 
energy $E_p \approx E_p^{pk}$, with more than half of 
this energy released in the form of leptons and photons which generates an electromagnetic cascade 
emerging in the form of $\gamma$ rays when the system becomes optically thin to $\gamma\gamma$ processes.
At $E_p \gg E_p^{pk}$, a larger fraction is extracted.
Consequently, the energy of baryons must be $\approx 10$ to 100 $\times$ greater than the nonthermal lepton content
if a comparable amount of electromagnetic radiation is to be emitted from hadronic as leptonic processes.
 Note the sensitive $\Gamma$-dependence of 
the photohadronic efficiency, $\eta_{p\gamma}(E^{pk}_p)\propto \Gamma^{-4}$, by comparison 
with $\tau_{\gamma\gamma}(\e) \psim \Gamma^{-6}$. The implications for neutrino and $\gamma$-ray 
production are considered in \cite{rmz04}. 

\section{Summary}

In this contribution, we have sketched the energy requirements for GRBs to be sources 
of UHECRs, and for electromagnetic signatures of ultrarelativistic hadrons to be found 
in the {\it Fermi} data. The large amounts of energy needed has been noted many times
in the past, whether from proton synchrotron \cite{tot98,zm01}  or photohadronic 
processes. Although the large energy requirements make uncomfortable demands on $\gamma$-ray 
emission models, an internal 
consistency is found insofar as the baryon load in long GRBs must be large given 
their relative rarity within the GZK radius, and that enormous energies are available from 
the rotational and accretion energy in the newly forming black holes. Future {\it Fermi} observations
and  the possibility of detecting PeV neutrinos from GRBs with IceCube could establish
whether GRBs are the sources of UHECRs.

\vskip0.2in

\noindent {\bf Acknowledgments:} I would like to thank Guido Chincarini and Bing Zhang for their kind invitation
to the Venice Shocking Universe conference, Soeb Razzaque for discussions, and 
Justin Finke for a careful reading of the manuscript.
This work is supported by the Office of Naval Research.

%



%
\end{document}